\documentclass[10pt]{{amsart}}
\usepackage{amssymb,amsfonts,latexsym,amscd}
\usepackage[dvips]{graphicx} 

\newcommand{\R}{\mathbb{R}}

\newcommand{\dst}{{\rm dist\,}}

\newtheorem{Theorem}{Theorem}[section]
\newtheorem{Lemma}{Lemma}[section]

\theoremstyle{remark}

\author{Rafael D. Benguria, Michael Loss and Heinz Siedentop}

\email{rbenguri@fis.puc.cl, loss@math.gatech.edu, h.s@lmu.de}
\title[Stability of an UTFW model]
{Stability of atoms and molecules in an ultrarelativistic
Thomas--Fermi--Weizs\"acker model}
\address{Department of Physics, Pontificia Universidad Cat\'olica de Chile Casilla 306, Correo 22
Santiago, Chile} \address{School of Mathematics, Georgia Institute
of Technology, Atlanta, GA 30332} \address{Mathematisches Institut,
Ludwig--Maximilians--Universit\"at M\"unchen, Theresienstrasse 39,
80333 M\"unchen, Germany}
\thanks{R.B. was supported by FONDECYT (Chile) projects \# 706--0200 and \#
106--0651 and CONICYT (Chile) PBCT Proyecto Anillo de
Investigaci\'on en Ciencia y Tecnolog\'\i a ACT30/2006; M.L. was
supported by NSF grant DMS--0600037; H.S. was supported by the
Deutsche Forschungsgemeinschaft grant SI 348/13-1}

\thanks{\copyright{2007 by the authors. This paper may be reproduced,
in its entirety, for non--commercial purposes}}

\begin{document}

\begin{abstract}
We consider the zero mass limit of a relativistic
Thomas--Fermi--Weizs\"acker model of atoms and molecules. We find
bounds for the critical nuclear charges that insure  stability.

\end{abstract}
\maketitle

\section{Introduction} \label{SectionIntroduction}
The zero mass limit of the relativistic Thomas--Fermi--Weizs\"acker
(henceforth ultrarelativistic TFW) energy functional for nuclei of
charges $z_i>0$ (which need not be integral) located at $R_i$, $i=1,
\dots, K$ is defined by \cite{ED87,ED88}
\begin{equation}
\xi(\rho)=a^{2} \int(\nabla \rho^{1/3} )^{2} \, dx + b^{2} \int
\rho^{4/3} \, dx -\int V(x){\rho}(x)\, dx + D(\rho,\rho),
\label{eq:a1}
\end{equation}
Here $\rho(x) \ge 0$ is the electron density,
\begin{equation}
V(x)= \alpha \sum_{i=1}^{K} \frac{z_i}{\vert x - R_i\vert}.
\label{eq:a2}
\end{equation}
the electrostatic potential created by the nuclei,
\begin{equation}
D(\rho,\rho)=\frac{\alpha}{2}\int \frac{\rho(x)\rho(y)}{|x-y|} \, dx \,  dy,
\label{eq:a3}
\end{equation}
the electronic repulsion, and $\alpha = e^2/\hbar c \approx 1/137$
is the fine structure constant. In units in which $\hbar=c=1$, the
constants $a^2$, and $b^2$ in (\ref{eq:a1}) are given respectively
by
\begin{equation}
a^2=\frac{3}{8\pi^2} (3\pi^2)^{2/3} \lambda
\label{eq:a4}
\end{equation}
and
\begin{equation}
b^2=\frac{3}{4}(3\pi^2)^{1/3}.
\label{eq:a5}
\end{equation}
In the non relativistic case, some emphasis has been placed on the
question of an appropriate choice of the coefficient $\lambda$ of
the gradient term connected to the kinetic energy. The non
relativistic gradient correction $\lambda (\nabla \rho)^2/\rho$ was
initially derived by Weizs\"acker \cite{VW35} with a value
$\lambda=1$, whereas the systematic gradient expansion by Kirznits
\cite{Ki67, Ho73} leads to $\lambda=1/9$. In the derivation of
Tomishima and Yonei \cite{TY66}, $\lambda=1/5$. Lieb
\cite{EL81,EL82} showed, that the Weizs\"acker term introduces a
$z^2$ correction to the leading $z^{7/3}$ term of the
non-relativistic ground state energy. Adapting the coefficient in
front of the Weizs\"acker term such that the correction agrees with
the leading $z^2$ correction of the non-relativistic quantum
mechanics, the Scott correction, lead Lieb to propose
$\lambda=0.185$. In relativistic quantum mechanics, the leading
energy correction remains unchanged whereas the Scott correction is
smaller than in relativistic quantum mechanics \cite{FSW07}, i.e.,
we cannot expect to have $\lambda$ the same value as in the TFW
functional. However, we are not yet in a position to proceed with
Lieb's strategy and to infer the coefficient of the gradient
correction from this. In particular that would require showing that
the massive equivalent of the function $\xi$ leaves indeed the
leading energy contribution unchanged and the gradient term yields
again a $z^2$ correction.

Let us first consider the atomic case, i.e., the case $K=1$,
$z_1=z$, $R_1=0$. Because of simple scaling considerations, if we
minimize the energy functional (\ref{eq:a1}) over all functions
$\rho$ for which each of the terms in (\ref{eq:a1}) makes sense, the
infimum of the energy functional is either zero or minus infinity.
In the first case we say the atom is {\it stable}. Otherwise we say
the atom is {\it unstable}. Our purpose here is to determine the
range of values of the  $z_i$'s for which the atom or molecule is
{\it stable}. The following result holds in the atomic case (i.e.,
for $K=1$, $z_1=z$, $R_1=0$) \cite{BePO02}.

\begin{Theorem}\label{thm:t1}
Let
\begin{equation}
\xi(\rho)=a^{2} \int(\nabla \rho^{1/3} )^{2} dx + b^{2} \int \rho^{4/3} dx
-\int z \alpha \frac{\rho}{|x|} dx + D(\rho,\rho),
\label{eq:5b}
\end{equation}
with $D(\rho,\rho)$ given by (\ref{eq:a3}).
Then
\begin{equation}
\inf \xi(\rho) = \left\{
\begin{array}{ll}
-\infty &  \mbox{for} \qquad z > \frac{4ab}{3\alpha} + \frac{7 \pi a^{3}}{6 b^{3}}
\\
0 &  \mbox{for} \qquad z <  \frac{4ab}{3\alpha}
\end{array} \right.
\label{eq:a6}
\end{equation}
where the infimum is taken over all nonnegative functions $\rho(x)$, such that
$\rho \in L^{4/3}(\R^3)$, $\nabla \rho^{1/3}\in L^2(\R^3)$, and $D(\rho,\rho)<\infty$.
\end{Theorem}

\bigskip
\noindent{\it Remarks}

\noindent i) It follows from (\ref{eq:a6}) that if $z<4ab/(3\alpha)$
the atom is stable, whereas if $z>4ab/(3\alpha) + 7\pi a^3/(6b^3)$
the atom is unstable. The exact critical value of $z$ ($z_c$ say)
dividing the region of stability from the region of unstability is
not known. However, it turns out that for the physical values of the
constants the gap between the upper and lower bounds on $z_c$ is
less than one, and therefore negligible (see the following remarks).

\noindent ii) For the physical values of $a$ and $b$ given by
(\ref{eq:a4}) and (\ref{eq:a5}), the atom will be stable if $z<
\sqrt{3 \lambda/2}/\alpha \approx 167.8 \sqrt{\lambda}$. Thus, if
$\lambda=1/9$ (i.e., the value used by  Kirznits, \cite{Ki67, Ho73})
the atom is stable if $z<56$. If $\lambda=1/5$ (i.e., the value used
by Tomishima and Yonei, \cite{TY66}) the atom is stable if $z < 75$.
Finally, using the value of Lieb \cite{EL81,EL82}, the atom is
stable if $z < 73$.

\noindent iii) As for the value of the gap, using the physical
values of the constants, one gets $7\pi a^3/(3b^6)=(7/12\pi)\sqrt{3
\lambda^3/2} < 0.021 < 1$  for all the values of $\lambda$
considered above.  Thus, the gap is negligible from the physical
point of view.

\bigskip

For the molecular case, i.e., when $K>1$, and $V$ is given by
(\ref{eq:a2}) the following result was proven in \cite{BePO02}.

\begin{Theorem}\label{thm:t2}
Let
\begin{equation}
\xi(\rho)=a^{2} \int(\nabla \rho^{1/3} )^{2} dx + b^{2} \int \rho^{4/3} dx
-\int V \,\rho \, dx + D(\rho,\rho),
\label{eq:c8}
\end{equation}
with $V$ given by (\ref{eq:a2}) and $D(\rho,\rho)$ given by (\ref{eq:a3}).
Then
\begin{equation}
\inf \xi(\rho) = 0 \qquad \mbox{if $Z=\sum_{i=1}^{K} z_i \le
\frac{4ab}{3 \alpha}$}, \label{eq:c9}
\end{equation}
where the infimum is taken over all nonnegative functions $\rho(x)$, such that $\rho \in L^{4/3}(\R^3)$,
$\nabla \rho^{1/3}\in L^2(\R^3)$, and $D(\rho,\rho)<\infty$.
\end{Theorem}

The above result is just a trivial extension of the atomic to the
molecular case, and in some sense is the best possible when the
interaction between the nuclei is not taken into account. In fact,
if we neglect the nuclear interaction one can always think of the
possibility of putting all the nuclear charges at the same point and
reducing the molecular case to the atomic case, which explains the
condition (\ref{eq:c9}) on $Z\equiv \sum_{i=1}^K z_i$. The
deficiencies of the above result are obvious. The goal of this paper
is to have a result that yields stability for reasonable values of
the nuclear charges. For that purpose the nucleus--nucleus
interaction
\begin{equation}
U \equiv  \alpha \sum_{1 \le i < j \le K} \frac{z_i \,
z_j}{|R_i-R_j|}, \label{eq:U}
\end{equation}
plays a key role, because it prevents the possibility of putting the
nuclear charges on top of each other.

Our main result is the following theorem for the molecular case.
\begin{Theorem}\label{thm:t3}
Let $\xi(\rho)$ be given by (\ref{eq:c8}) for functions $\rho$ as in
Theorem \ref{thm:t2}, with $V$ given by (\ref{eq:a2}). Then, we have
stability, i.e.,
$$
\inf \xi(\rho) + U \ge 0
$$
provided
\begin{equation}
0 \le z_i \le \frac{4a}{3 \alpha} b \sqrt{1-x} \label{eq:c10}
\end{equation}
where $x \in (0,1)$ is the root of
\begin{equation}
\frac{1-x}{x^3}=\frac{b^4}{a^2} \left(\frac{4}{3} \right)^2
\frac{1}{2 \pi \alpha(4+9\alpha^4)}. \label{eq:c11}
\end{equation}
\end{Theorem}

\bigskip
\noindent{\it Remark} For the physical values of $a$ and $b$ given
by (\ref{eq:a4}) and (\ref{eq:a5}), and taken the physical value of
the fine structure constant (i.e., $\alpha=1/137$) in
(\ref{eq:c11}), Theorem \ref{thm:t3} says that the molecule will be
stable if each $z_i \le 55$, if $\lambda=1/9$ (i.e., the value used
by Kirznits, \cite{Ki67, Ho73}). If $\lambda=1/5$ (i.e., the value
used by Tomishima and Yonei, \cite{TY66}) the molecule is stable if
each $z_i\le 74$. Finally, using the value of Lieb \cite{EL81,EL82},
the molecule is stable if each $z_i \le  71$. These bounds on each
individual nuclear charge are almost the same as those embodied in
the atomic case (i.e., the ones given in Theorem \ref{thm:t1}
above).

\bigskip

In the next section we give the proof of Theorem \ref{thm:t3}.

\section{Improved results on the stability of molecules}

In this section we prove Theorem \ref{thm:t3}. Our proof relies in a
{\it modified uncertainty principle} (see Theorem \ref{thm:td1}
below) which is of independent interest. As we mentioned in the
introduction, the nucleus--nucleus interaction plays a key role in
the stability of molecules. As in \cite{Lieb-Daubechies} we may use
the fact that the energy is separately concave in the nuclear
charges and each charge $z_i$ varies between $0$ and $z$. The
minimum of a concave function is always on the boundary and hence
the value od $z_i$ wants to be either $0$ or $z$. If it is zero we
have one nucleus less and if it is $z$ then we are in the case we
are considering.

First we need some notation. We introduce the nearest neighbor, or Voronoi, cells
\cite{Vo07} $\{ \Gamma_j \}_{j=1}^K$ defined by
\begin{equation}
\Gamma_j = \{ x \bigm| |x-R_j| \le |x-R_k|\}. \label{eq:v1}
\end{equation}

The boundary of $\Gamma_j$, $\partial \Gamma_j$, consists of a finite number of planes.
We also define the distance
\begin{equation}
D_j = \dst(R_j,\partial \Gamma_j) = \frac{1}{2} \min \{ |R_k-R_j| \bigm| j\neq k \}.
\label{eq:v2}
\end{equation}

One of the key ingredients we need in the sequel is an electrostatic
inequality of Lieb and Yau \cite{LYI88, LYII88}. Define a function
$\Phi$ on $\R^3$ with the aid of the Voronoi cells mentioned above.
In the cell $\Gamma_j$, $\Phi$ equals the electrostatic potential
generated by all the nuclei except for the nucleus situated in
$\Gamma_j$ itself, i.e., for $x \in \Gamma_j$,
\begin{equation}
\Phi(x) \equiv z \sum_{\stackrel{i=1}{i \neq j}}^K |x-R_i|^{-1}.
\label{eq:2.1}
\end{equation}
If $\nu$ is any bounded Borel measure on $\R^3$ (not necessarily positive) then
\begin{equation}
\frac{1}{2} \int_{\R^3} \int_{\R^3} |x-y|^{-1} \, d \nu (x) \, d \nu (y)
- \int_{\R^3} \Phi(x) \, d \nu (x) + U
\ge \frac{1}{8} z^2 \sum_{j=1}^{K} D_j^{-1}.
\label{eq:2.2}
\end{equation}

We will also need a {\it localization} result for the kinetic energy which will allow us to
control the Coulomb potential near each nuclei. This localization result for the UTFW model
is given by Theorem  \ref{thm:td1} below.

\bigskip
\begin{Theorem}[Modified uncertainty principle] \label{thm:td1}
For any smooth function $f$ on the closed ball $B_R$ of radius $R$
we have the estimate
$$
a^2 \int_{B_R} |\nabla f(x)|^2 dx + b^2 \int_{B_R} f(x)^4 dx \ge ab
\int_{B_R} \left[{4 \over 3|x|}-{2 \over R}\right] f(x)^3 dx.
$$
\end{Theorem}
\bigskip
\noindent{\it Remark} Notice that the factor $4/3$ is best possible
and it agrees with the sharp value given in Theorem 1.1.

\bigskip
\noindent To prove this theorem, we need the following preliminary
result.
\begin{Lemma}
Let $u(r)$ be any smooth function with $u(R)=0$. Then
the following uncertainty principle holds
$$
|\int_{B_R} (3u(|x|)+ u'(|x|)|x|) f(x)^3 dx|
\le 3 (\int_{B_R} |\nabla f(x)|^2 dx)^{1/2} (\int_{B_R} u(|x|)^2 |x|^2  f(x)^4 dx)^{1/2} \ .
$$
There is equality if and only if
$$
f = \frac{1}{\lambda \int_0^r[su(s)]ds +C}
$$
for some constants  $C$ and $\lambda$.
\end{Lemma}
\begin{proof}
Set
$$
g_i(x) = u(|x|)x_i
$$
where $u$ is a smooth function with $u(R) = 0$ and note that
$$
\int_{B_R} (3u(|x|)+ u'(|x|)|x|) f(x)^3 dx =\sum_j \int_{B_R}f(x)
[\partial_j \, g_j(x)] f(x)^2 dx
$$
$$
= \sum_j \int_{B_R} f(x) \partial_j(g_jf^2)(x) dx
-2 \sum_j \int_{B_R}f(x)^2 g_j(x) \partial_j f(x) dx \ .
$$
Integrating the first term by parts yields
$$
\sum_j \int_{B_R} f(x) \partial_j(g_jf^2)(x) dx =-\sum_j \int_{B_R}
\partial_j f(x) g_j(x) f(x)^2 dx + \int_{\partial B_R} f(x)^3
u(|x|)|x| dS(x),
$$
where the boundary term vanishes since $u(R)=0$.
Thus,
$$
\int_{B_R} (3u(|x|)+ u'(|x|)|x|) f(x)^3 dx = -3 \sum_j
\int_{B_R}f(x)^2 g_j(x) \partial_j f(x) dx.
$$
Using Schwarz' inequality on the last term yields
$$
\left|\int_{B_R} (3u(|x|)+ u'(|x|)|x|) f(x)^3 dx\right| \le 3
\left(\int_{B_R} |\nabla f(x)|^2 dx\right)^{1/2} \left(\int_{B_R}
u(|x|)^2 |x|^2 f(x)^4 dx\right)^{1/2} \ .
$$
Schwarz's inequality is an equality if and only if
$$
\partial_jf = -\lambda g_j(x) f(x)^2 \ ,
$$
which can easily be integrated and yields the stated function.
\end{proof}

\bigskip
\begin{proof} [Proof of Theorem \ref{thm:td1}]
To prove the theorem, pick
$$
u(r) = {1 \over 2}({1 \over r}-{1 \over R})
$$
in the lemma which
leads to the inequality
$$
\int_{B_R}\left[{1 \over |x|} -{3 \over 2R}\right] f(x)^3 dx \le {3
\over 2} \left(\int_{B_R} |\nabla f(x)|^2 dx\right)^{1/2}
\left(\int_{B_R}(1-{|x| \over R})^2 f(x)^4 dx\right)^{1/2}
$$
Next, applying the inequality between the arithmetic and geometric mean yields
$$
a^2 \int_{B_R} |\nabla f(x)|^2 dx + b^2 \int_{B_R} \left(1-{|x|
\over R}\right)^2 f(x)^4 dx \ge ab \int_{B_R} \left[{4 \over 3|x|}-
{2 \over R}\right] f(x)^3 dx \ ,
$$
from which the theorem follows.
\end{proof}
\bigskip

In order to prove our main result, we consider the total energy,
$\xi(\rho) + U$, and we split the $\int \rho^{4/3}$ term in two
parts, i.e., $(b_1^2+b_2^2)\int \rho^{4/3}$. For later discussions,
it is important to remark that the parameter $b_2$ can be chosen
arbitrarily in the interval $(0,b)$. Then we use Theorem
\ref{thm:td1} with $f^3=\rho$ and  $B_R=B_j$, the largest ball
inscribed in the corresponding Voronoi cell $\Gamma_j$,  to get
\begin{eqnarray}
\xi(\rho) + U   \ge & b_1^2 \int_{\R^3} \rho^{4/3} \, dx - \int_{\R^3} V(x) \rho (x) \nonumber
\\
+ & a b_2 \sum_{j=1}^K
\int_{B_j} \left( \frac{4}{3 \vert x - R_j \vert} - \frac{2}{D_j} \right)
\rho(x) \, dx + D(\rho,\rho) + U.
\label{eq:d12}
\end{eqnarray}
In order to cancel the Coulomb singularity inside $B_j$ we choose
the parameter
\begin{equation}
b_2 = \frac{3}{4} \frac{\alpha \, z}{a}. \label{eq:b2-z}
\end{equation}
The restrictions on $b_2$ will give restrictions on
$z$ to insure stability.

With the help of the Voronoi cells we now define,
\begin{equation}
W(x) \equiv \Phi (x) + \frac{z}{|x-R_j|}
\label{eq:d13}
\end{equation}
if $x \in \Gamma_j$ and $|x-R_j| \ge  \, D_j$, whereas,
\begin{equation}
W(x) \equiv  \Phi(x) + \frac{2 a b_2}{D_j}, \label{eq:d14}
\end{equation}
if $x \in \Gamma_j$ and $|x-R_j| \le D_j$.
Now, if we restrict to values of $z$ such that  $z \le 4 a b_2 /(3 \alpha)$, we can finally write
\begin{equation}
\xi(\rho) + U \ge \xi_1(\rho) + \xi_2(\rho),
\label{eq:d15}
\end{equation}
with
\begin{equation}
\xi_1(\rho) = b_1^2 \int_{\R^3} \rho^{4/3} \, dx - \alpha \int_{\R^3} W(x) \rho(x) + \alpha \int_{\R^3} \Phi \, \rho(x) \, dx,
\label{eq:d16}
\end{equation}
and
\begin{equation}
\xi_2(\rho) =  D(\rho,\rho) - \alpha \int_{\R^3} \Phi(x) \rho(x) \, dx +  U
\label{eq:d17}
\end{equation}

Using the Lieb--Yau electrostatic inequality (\ref{eq:2.2}) we have
\begin{equation}
\xi_2 (\rho) \ge \alpha \frac{z^2}{8} \sum_{j=1}^K \frac{1}{D_j}.
\label{eq:d18}
\end{equation}

On the other hand, it is simple to estimate $\xi_1(\rho)$ from below, since
it is simple to solve the variational principle $\inf_{\rho} \xi_1(\rho)$.
Thus we get,
\begin{equation}
\xi_1(\rho) \ge -\frac{1}{4} \alpha^4 \left( \frac{3}{4 b_1^2} \right)^3 \int_{\R^3} (W-\Phi)_+^4 \, dx.
\label{eq:d19}
\end{equation}
From (\ref{eq:d15}), (\ref{eq:d18}), and (\ref{eq:d19}) we get
\begin{equation}
\xi(\rho) + U \ge  -\frac{1}{4} \alpha^4 \left( \frac{3}{4 b_1^2} \right)^3 \int_{\R^3} (W-\Phi)_+^4 \, dx+
\alpha \frac{z^2}{8} \sum_{j=1}^K \frac{1}{D_j}.
\label{eq:d20}
\end{equation}
Now, using (\ref{eq:d14}) we compute,
\begin{equation}
\int_{B_j} (W-\Phi)_+^4 \, dx = \left( \frac{2ab_2}{D_j} \right)^4 \frac{4}{3} \pi D_j^3 =
\frac{64 \pi a^4 b_2^4}{3 D_j}.
\label{eq:d21}
\end{equation}
Since any Voronoi cell is contained in a half space, one calculates
as in \cite{LLS}  that
\begin{equation}
\int_{\Gamma_j \setminus B_j} (W-\Phi)_+^4 \, dx = \int_{\Gamma_j
\setminus B_j} \left( \frac{z}{\vert x-R_j \vert}\right)^4 \, dx \le
\frac{3 \pi z^4}{D_j}. \label{eq:d22}
\end{equation}

Finally, using the estimates (\ref{eq:d21}) and (\ref{eq:d22}) in (\ref{eq:d20}), we get
\begin{equation}
\xi(\rho)+ U \ge \frac{\alpha}{8} M \sum_{j=1}^K \frac{1}{D_j},
\label{eq:d23}
\end{equation}
where
\begin{equation}
M=  -\frac{1}{4} \alpha^4 \left( \frac{3}{4 b_1^2} \right)^3
\left( 3 \pi z^4 + \frac{64 \pi a^4 b_2^4}{3}\right) +\frac{\alpha z^2}{8} \ .
\label{eq:d24}
\end{equation}
If $M>0$ then $\xi(\rho) + U >0$ and the molecule is stable. Using
the fact that $b_2^2=b^2-b_1^2$ together with (\ref{eq:b2-z}) the
condition $M \ge 0$ can be written solely in term of $b_1$ as
\begin{equation}
a^2\frac{b^2-b_1^2}{b_1^6} \le \left(\frac{4}{3}\right)^2
\frac{1}{2\pi\left(4 \alpha + 9 \alpha^5\right)} \ .
\label{eq:d25}
\end{equation}
Finally, in order to allow for the largest possible value of $z$
(equivalently the largest possible value of $b_2$), $b_1$ must be
chosen so that we have equality in (\ref{eq:d25}). Since the left
side of (\ref{eq:d25}) is decreasing as a function of $b_1$ in the
allowed interval $(0,b)$, we conclude the proof of our Theorem
\ref{thm:t3}.

\bigskip

\noindent{\it Remarks}

\noindent i) Note that we do not need a bound on the fine structure
constant to ensure stability in this model. This is due to the
absence of the exchange term.

\noindent ii) If we set $\alpha =\frac{1}{137}$ we find stability up
to $z=71$, when the parameter $\lambda =0.185$, up to $z=74$ when
$\lambda = 0.2$ and $z=55$ when $\lambda = 1/9$.

\bigskip

\end{document}